# Alternative equation of motion approach to the single-impurity Anderson model

Grzegorz Górski, Jerzy Mizia and Krzysztof Kucab

Faculty of Mathematics and Natural Sciences,

University of Rzeszów, ul. Rejtana 16A, 35-959 Rzeszów, Poland

Solving the single-impurity Anderson model (SIAM) is a basic problem of solid state physics. The SIAM model is very important, at present it is also used for systems with quantum impurities, e.g. semiconductor quantum dots and molecular transistors. Its main application is in the scheme of dynamical mean field theory (DMFT) describing strong correlation electron systems. To solve the SIAM problem we use the equation of motion (EOM) Green function approach. In this report we present the novel EOM approximation in which we differentiate the Green function over both time variables. This differs from the commonly used EOM solution by Appelbaum, Penn and Lacroix where the authors take time derivative only over primary time variable. After extending calculations to higher order Green functions we find the new approximate dynamical solution of SIAM. The results are compared with the solutions to the SIAM problem at intermediate Coulomb repulsion $U$ such as the Modified Iterative Perturbation Theory. Our approach is suitable for describing quantum dots.

## 1. Introduction

The single-impurity Anderson model (SIAM) [1] quantizes the relations between the kinetic-energy (itinerant) and potential-energy (localized) effects. It describes systems with both localized and itinerant electrons. It was used to analyze mixed systems with valence and heavy fermion electrons. In recent years there has been a strong interest in the application of SIAM model to strongly correlated electrons since in the limit $d = \infty$ we can map the Hubbard model to the SIAM problem. This operation is based on the scheme of dynamical mean-field theory (DMFT) [2] describing well the dynamics of systems with strong correlations. Another field where the SIAM model is used is with the transport properties of nanoscale materials with quantum dots (QD) and single electron transistors. The SIAM theory predicts an enhancement of the dot conductance at low temperatures due to the development of the so-called Kondo resonance.

For non-interacting electrons ($U = 0$) the SIAM problem can be solved precisely. On the other hand, solving the SIAM model within the DMFT and for QD for systems with interaction ($U \neq 0$) requires the use of numerical methods (e.g. quantum Monte Carlo, exact diagonalization, or numerical renormalization group) or approximate analytical methods (e.g. modified iterative perturbation theory) (MPT) [3-5]). All of them have some limitations.

One of analytical methods for solving the SIAM problem is the equation-of-motion (EOM) approach [6,7]. Until now the deficiency of the EOM approach was not having the Fermi-liquid state at half filling for $U > 0$. The reason for this fault was that the EOM schema used strong coupling Green function expansion, which was losing the metallic effects, particularly in the particle-hole symmetric case. In this report we present a novel EOM approach in which we calculate the single particle Green function (and the density of states (DOS)) differentiating Green functions over both time variables. This differs from the commonly used EOM solution by Appelbaum, Penn [7] and Lacroix [6] where the authors take time derivative only over primary time variable. The results are comparable with the MPT which is an interpolative extension of the second-order perturbation theory. Our approach can be applied to analysis of the quantum dots.

## 2. Method

The SIAM model is based on the assumption that in the sea of conduction electrons (with energy dispersion $\varepsilon_k$) there is a localized impurity electron (with energy $\varepsilon_d$) for which we include the Coulomb interaction. The Hamiltonian of this model has the form

$$H = \sum_\sigma \varepsilon_d \hat{n}_{d\sigma} + \frac{U}{2}\sum_\sigma \hat{n}_{d\sigma}\hat{n}_{d-\sigma} + \sum_{k\sigma}(\varepsilon_k - \mu)\hat{n}_{k\sigma} + \sum_{k\sigma}\left(V_{dk}d_\sigma^+ c_{k\sigma} + h.c.\right), \quad (1)$$

where $d_\sigma^+ (d_\sigma)$ are the creation (annihilation) operators for the impurity electron, $c_{k\sigma}^+ (c_{k\sigma})$ are the creation (annihilation) operators for the conduction electron (bath), $U$ is the on-site Coulomb interaction between electrons on the impurity, and $V_{dk}$ is the coupling between the bath and impurity orbital ( we assume $V_{dk} = V$ ) . We will analyze the paramagnetic case, therefore the spin indices for energy ($\varepsilon_d$, $\varepsilon_k$, $\mu$) and interaction $V_{dk}$ will be neglected.

In our analysis we will use the equation of motion for the Green function method to solve the SIAM model. In general the EOM may be written as

$$\varepsilon \langle\langle A;B \rangle\rangle_\varepsilon = \langle [A,B]_+ \rangle + \langle\langle [A,H]_-;B \rangle\rangle_\varepsilon . \quad (2)$$

Applying it to the function $G_{d\sigma}(\varepsilon) = \langle\langle d_\sigma;d_\sigma^+ \rangle\rangle_\varepsilon$ we obtain:

$$[\varepsilon - \varepsilon_d - \Delta_\sigma(\varepsilon)]\langle\langle d_\sigma;d_\sigma^+ \rangle\rangle_\varepsilon = 1 + U\langle\langle \hat{n}_{d-\sigma}d_\sigma;d_\sigma^+ \rangle\rangle_\varepsilon , \quad (3)$$

where $\Delta_\sigma(\varepsilon) = \sum_k \frac{V^2}{\varepsilon + \mu - \varepsilon_k}$ is the hybridization function.

To solve eq. (3) we have to write the EOM for higher order Green function $\langle\langle \hat{n}_{d-\sigma} d_\sigma; d_\sigma^+ \rangle\rangle_\varepsilon$. Instead of eq. (2) we will use the method with differentiating over the second time ($t'$), which gives the EOM in the following form [8]:

$$-\varepsilon\langle\langle A; B\rangle\rangle_\varepsilon = -\langle[A,B]_+\rangle + \langle\langle A;[B,H]_-\rangle\rangle_\varepsilon \quad , \tag{4}$$

from which we obtain

$$[\varepsilon - \varepsilon_d - \Delta_\sigma(\varepsilon) - U]\langle\langle \hat{n}_{d-\sigma} d_\sigma; d_\sigma^+\rangle\rangle_\varepsilon = n_{d-\sigma} - U\langle\langle \hat{n}_{d-\sigma} d_\sigma; d_{-\sigma} d_{-\sigma}^+ d_\sigma^+\rangle\rangle_\varepsilon. \tag{5}$$

We will calculate function $\langle\langle \hat{n}_{d-\sigma} d_\sigma; d_{-\sigma} d_{-\sigma}^+ d_\sigma^+\rangle\rangle_\varepsilon$ using the Kuzemsky Green function decoupling (see [8] and [9]). Using these approximations we arrive at:

$$\langle\langle \hat{n}_{d-\sigma} d_\sigma; d_{-\sigma} d_{-\sigma}^+ d_\sigma^+\rangle\rangle_\varepsilon \approx n_{d-\sigma}(1 - n_{d-\sigma}) G_{d\sigma}(\varepsilon) - \Gamma_\sigma(\varepsilon, T) \quad , \tag{6}$$

$$\Gamma_\sigma(\varepsilon, T) = \iiint \frac{S_{d-\sigma}^{HF}(x) S_{d-\sigma}^{HF}(y) S_{d\sigma}^{HF}(z)}{\varepsilon + x - y - z + i0^+} F_T(T, x, y, z) dx\,dy\,dz \tag{7}$$

where

$$S_{d\sigma}^{HF}(\varepsilon) = -\frac{1}{\pi} \text{Im} \frac{1}{\varepsilon - \varepsilon_{d0} - \Delta_\sigma - Un_{-\sigma}} \quad , \tag{8}$$

$$F_T(T, x, y, z) = f(x)f(-y)f(-z) + f(-x)f(y)f(z) \quad . \tag{9}$$

Using Eqs. (5), (6) and (3) we can find the function:

$$G_{d\sigma}(\varepsilon) = \frac{1}{\varepsilon - \varepsilon_d + \mu - \Delta(\varepsilon) - \Sigma_{d\sigma}} \quad , \tag{10}$$

where the self-energy is given by:

$$\Sigma_{d\sigma}(\varepsilon) = Un_{d-\sigma} + \frac{U^2 \Gamma_{d\sigma}(\varepsilon)}{1 + U^2 \Gamma_{d\sigma}(\varepsilon) G_{d\sigma}^{HF}(\varepsilon)} \quad . \tag{11}$$

We analyzed the self-energy given by Eq. (11) in the DMFT scheme (see [9]). The results show that for large $U$ values the self-energy is not convergent. The reason for this is that the denominator $[1 + U^2 \Gamma_{d\sigma}(\varepsilon) G_\sigma^{HF}(\varepsilon)]^{-1}$ at larger $U$ has a singularity localized between the resonant peak and the

Hubbard bands. Now we will make an approximation which replaces the $G_\sigma^{HF}(\varepsilon)$ function by the parameter, being an average of $G_\sigma^{HF}(\varepsilon)$ functions at states $\varepsilon_d$ and $\varepsilon_d + U$ in the atomic limit

$$\alpha_\sigma = \frac{1}{2}\left[G_\sigma^{HF}(\varepsilon_d - \mu) + G_\sigma^{HF}(\varepsilon_d - \mu + U)\right]\bigg|_{V \to 0}$$
$$= \frac{\varepsilon_d - \mu - \varepsilon_{d0} + U/2(1 - 2n_{d-\sigma})}{(\varepsilon_d - \mu - \varepsilon_{d0})^2 + (\varepsilon_d - \mu - \varepsilon_{d0})U(1 - 2n_{d-\sigma}) - U^2 n_{d-\sigma}(1 - n_{d-\sigma})}, \quad (12)$$

which gives

$$\Sigma_{d\sigma}(\varepsilon) = U n_{d-\sigma} + \frac{U^2 \Gamma_{d\sigma}(\varepsilon)}{1 + \alpha U^2 \Gamma_{d\sigma}(\varepsilon)}. \quad (13)$$

Equation (13) is now quite similar to the equation for $\Sigma_{d\sigma}(\varepsilon)$ in the MPT theory [3].

## 3. Results and conclusions

In Fig. 1 we present the spectral density of states $S_{d\sigma}(\varepsilon) = -\frac{1}{\pi} \operatorname{Im} G_{d\sigma}(\varepsilon)$ for different interactions $V$ at the half-filling point. The DOS presented shows the three peaks structure with quasiparticle resonance peak at the Fermi energy, and two broad satellite sub-bands corresponding to the atomic quasiparticle levels $\varepsilon_d$ and $\varepsilon_d + U$. Decreasing the value of the hybridization parameter $V$, we observe narrowing of the resonance peak. The value of DOS at the Fermi level stays the same as in the non-interacting case.

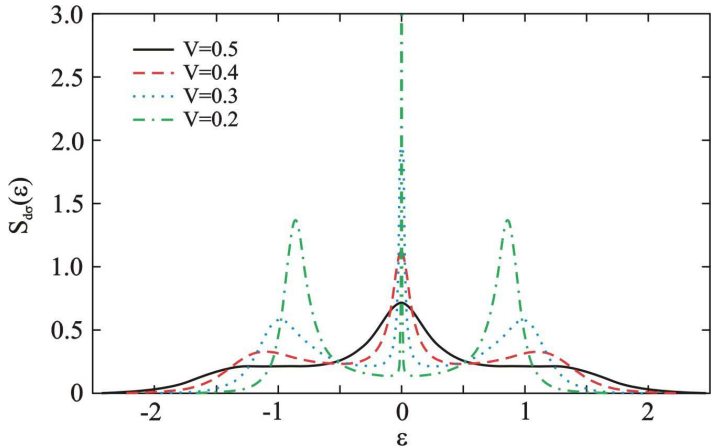

Fig. 1 Spectral density $S_{d\sigma}(\varepsilon)$ as a function of energy $\varepsilon$ calculated for different interactions $V$. The case of half-filled band, $U = 1.5$ and $T = 0$. The unit of energy is half bandwidth of non-interactive case.

In Fig. 2 below we present the spectral density of states for different *d*-electrons concentrations ($n_d$). For lower values of concentration $n_d$ the resonant level shifts towards lower energies and mixes with the energy level corresponding to energy $\varepsilon_d$. This results in significant broadening of the resonant level.

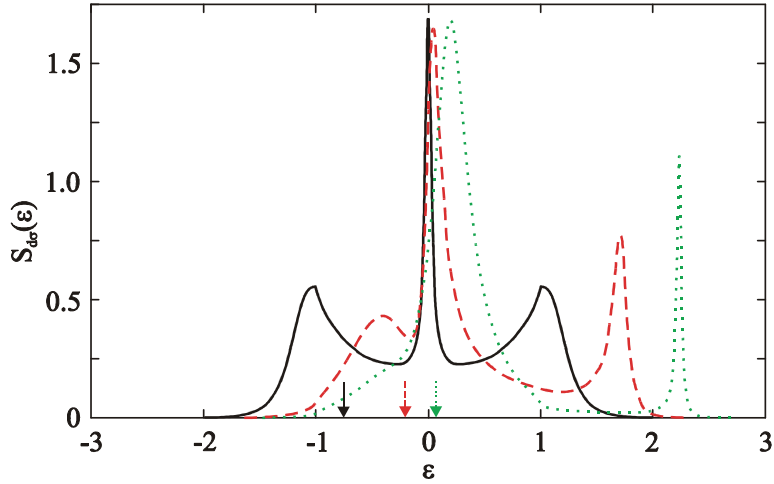

Fig. 2 Spectral density $S_{d\sigma}(\varepsilon)$ as a function of energy $\varepsilon$ calculated for different values of $n_d$ ($n_d = 1$ -solid black line, $n_d = 0.7$ -dashed red line, $n_d = 0.4$ - dotted green line), $U = 1.5$, $V = 0.3$, $\mu = 0$ and $T = 0$. The level $\varepsilon_d$ is marked for different concentrations by arrows. The unit of energy is half bandwidth of non-interactive case.

In Fig. 3 below we present the spectral density of states obtained under assumption of constant value of the hybridization function $\Delta_\sigma(\varepsilon) \equiv \Delta = i\Gamma$. This kind of approximation is characteristic in models describing the quantum dots physics. The resonant peak obtained is characterized by a larger of both the maximum value and the width with respect to results obtained by Lacroix approach [6].

In summary, using the EOM approach we obtained the DOS with the three peaks structure composed of quasiparticle resonance peak at the Fermi energy, and two broad satellite sub-bands corresponding to the atomic quasiparticle levels. These results obtained by our EOM method are consistent with the results obtained within the MPT method [3-5]. As opposed to the MPT method the key equation (13) is derived analytically and not introduced in the phenomenological way. The approach presented here can be used to describe quantum dots (see Fig. 3).

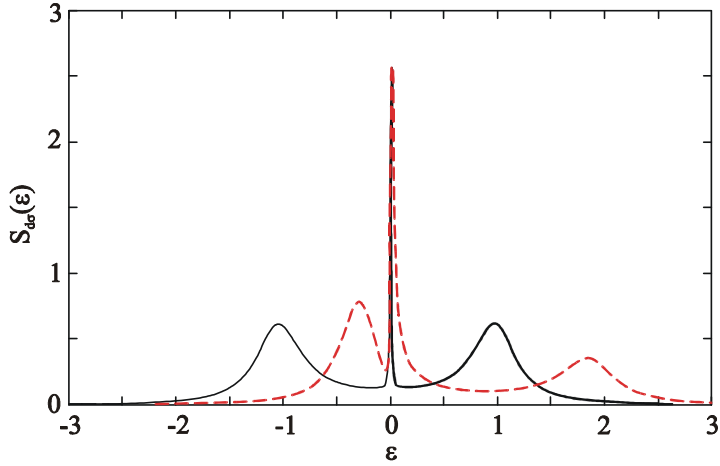

Fig. 3 Density of states of the QD in the equilibrium case for a finite Coulomb energy $U = 20\Gamma$, $\Delta = i\Gamma$, $\Gamma = 0.1$, $T = 0$ and different values of $n_d$. ($n_d = 1$ – solid black line, $n_d = 0.8$ – dashed red line). The unit of energy is half bandwidth of the lead.

Acknowledgments

We would like to thank Prof. W. Nolting and Prof. J. Barnaś for many valuable discussions and suggestions.

**References**

[1] P.W. Anderson, Phys. Rev. **124** (1), 41 (1961).

[2] A. Georges, G. Kotliar, W. Krauth and M. J. Rozenberg, Rev. Mod. Phys. **68**, 13 (1996).

[3] D. Meyer, T. Wegner, M. Potthoff, W. Nolting, Physica B **270,** 225 (1999).

[4] H. Kajueter and G. Kotliar, Phys. Rev. Lett. **77**, 131 (1996).

[5] M. Potthoff, T. Wegner, and W. Nolting, Phys. Rev. B **55**, 16132 (1997).

[6] C. Lacroix, J. Appl. Phys. **53**, 2131 (1982).

[7] J.A. Appelbaum and D.R. Penn, Phys. Rev. **188**, 874 (1969).

[8] A.L. Kuzemsky, Riv. Nuovo Cimento **25**, 1 (2002).

[9] G. Górski and J. Mizia, Physica B **427**, 42 (2013).